
\input harvmac.tex
%
%
\def\bp{{\beta_p}}
\def\frac#1#2{{#1\over#2}}
\def\half{\frac12}
\font\cmss=cmss10 \font\cmsss=cmss10 at 7pt
\def\IZ{\relax\ifmmode\mathchoice
{\hbox{\cmss Z\kern-.4em Z}}{\hbox{\cmss Z\kern-.4em Z}}
{\lower.9pt\hbox{\cmsss Z\kern-.4em Z}}
{\lower1.2pt\hbox{\cmsss Z\kern-.4em Z}}\else{\cmss Z\kern-.4em Z}\fi}
\def\CF{{\cal F}}
\def\CL{{\cal L}}
\def\CT{{\cal T}}
\def\CZ{{\cal Z}}
%
%
\lref\Zam{A. B. Zamolodchikov, Sov. J. Nucl. Phys. {\bf 44} (1986)
529.}
\lref\matmod{D. J. Gross and A. A. Migdal, Phys. Rev. Lett. {\bf 64}
(1990) 717; M. Douglas and S. Shenker, Nucl. Phys. {\bf B335} (1990)
635; E. Brezin and V. Kazakov, Phys. Lett. {\bf 236B} (1990) 144;
M. Douglas, Phys. Lett. {\bf 238B} (1990) 176; E. Brezin, V. Kazakov
and Al. Zamolodchikov, Nucl. Phys. {\bf B338} (1990) 673;
D. Gross and N. Miljkovic, Phys. Lett. {\bf 238B} (1990) 217;
P. Ginsparg and J. Zinn Justin, Phys. Lett. {\bf 240B} (1990) 333.}
\lref\GK{D. Gross and I. Klebanov, Nucl. Phys. {\bf B344} (1990) 639.}
\lref\Moo{G. Moore, Yale preprint YCTP-P1-92 (1992).}
\lref\Kut{D. Kutasov, Mod. Phys. Lett. {\bf A7}  (1992) 2943.}
\lref\BK{M. Bershadsky and I. R. Klebanov, Phys. Rev. Lett. {\bf 65} (1990)
3088.}
\lref\KL{I. R. Klebanov and D. Lowe, Nucl. Phys. {\bf B363} (1991) 543.}
\lref\BPZ{A. Belavin, A. Polyakov, and A. B. Zamolodchikov, Nucl. Phys.
{bf B241} (1984) 333.}
\Title{\vbox{\baselineskip12pt\hbox{PUPT-1357}
\hbox{hep-th/9212023}}}
{\vbox{\centerline{The Gravitational Sine-Gordon Model}}}
\centerline{{Edward Hsu}\footnote{$^\dagger$}
{E-mail address: \it{hsu@puhep1.princeton.edu}} and David Kutasov}
        \vskip2pt\centerline{\it{Department of Physics}}
        \vskip1pt\centerline{\it{Princeton University}}
        \vskip1pt\centerline{\it{Princeton, NJ 08544 USA}}
\rm
\vskip .5in
\noindent
We use matrix model results to investigate the Sine-Gordon model coupled to
two dimensional gravity.  For relevant (in the RG sense)
potentials, we show that the
$c=1$ string, which appears in the ultraviolet limit of this model,
flows to a set of decoupled $c=0$ (pure gravity) models in the infrared.
The torus partition sum, which was
argued previously to count the number of string degrees of
freedom and hence satisfy a new $c$ -- theorem, is shown to be a
monotonically decreasing function of the scale (given by the quantum
area of the world-sheet).
The model discussed describes an interesting time dependent solution of
two dimensional string theory.

\Date{December 1992}

\newsec{Introduction}

The Lagrangian
\eqn\a{
\CL=(\partial X)^2+\sum_{i=1}^N \lambda _i \cos{p_i X}
}
for an interacting scalar field in two dimensions generalizes
the Sine-Gordon (SG) model (the case with $N=1$).  In general not much is
known about the Quantum Field Theory (QFT)
to which it gives rise, but a qualitative picture of
what one may expect arises from the work of Zamolodchikov \Zam, according
to which the infrared limit is governed by the minima of the effective
potential for $X$, $V(X)$.  According to this picture, at large scales
the $X$ field freezes at one of these minima and is described by a
minimal model; precisely which model depends on the behavior of $V(X)$
around the minimum.  If $V(X)$ has more than one minimum, the theory
presumably splits into a decoupled set of minimal models.

While this picture is plausible and has received a certain amount of
verification over the years, exact calculations in \a~are difficult
because of the large fluctuations of the $X$ field in two dimensions.
For example, the effective potential $V(X)$ is renormalized and may
differ significantly from $\sum_{i=1}^N \lambda _i \cos{p_i X}$ in \a.

{}From this point of view, it is interesting to consider the theory
\a~coupled to gravity.  Two dimensional gravity coupled to a scalar field,
and the
generalization \a, is quite well understood \matmod
\ref\moore{G. Moore, Nucl. Phys. {\bf B368} (1992) 557;
G. Moore, R. Plesser and S. Ramgoolam, Nucl. Phys. {\bf B377} (1992)
143.};
thus one may hope to
develop a good understanding of the space of $\{\lambda _i\}$, at least
in the presence of gravity.  But then, it is generally expected that the
structure of the space of couplings is the same with or without gravity;
e.g., if \a~exhibits a renormalization group (RG) trajectory leading to
a certain CFT in the infrared limit in the presence of gravity, one would
expect this trajectory to exist also when gravity is turned off (the
converse is obviously true).  Hence, one may study the space of QFT's
\a~using the exact results of matrix models.
An example of what we have in mind is the space of minimal models.  In
flat space QFT only very partial results regarding this space
have been established; after coupling to gravity the problem is exactly
solvable in terms of certain integrable hierarchies \matmod.
In particular the
flow structure (at least genus by genus) is trivial to obtain and it of
course agrees with the known results in the absence of gravity.

An additional motivation to study the model \a~coupled to gravity stems
from the interpretation of this system as a string theory in two
dimensional space-time.  There are several applications for \a~in that
context.  One is to the study of time dependent solutions of string
theory.  Equation \a~describes a solution in which a non-trivial
``tachyon'' profile is propagating through the two dimensional space-time.
There have been many discussions of gravitational back-reaction to the
``tachyon'' stress tensor, but the issue is still not completely resolved.
Perhaps a better understanding of \a~will contribute to the solution of
that problem.
In addition, a good understanding of non-trivial time dependent vacua is
bound to be useful.

A second application of \a~is to the study of the role of non-singlet
states in the $c=1$ matrix model.  According to \GK~non-singlets
correspond in the continuum to vortices (winding modes).  Thus one
imagines having in the action terms like \a~(which in the dual picture
correspond to winding states), with $p_i=n_i R$, $R$ being the radius of
$X$.  At large enough radius all winding terms are irrelevant (in the RG
sense).  As the radius decreases, first one ($n=1$), then two ($n=1,2$),
etc. winding modes become relevant and \a~describes the resulting
theory.  One expects \GK~a dynamical transition to a theory where $X$ is
discrete, but the detailed picture should follow from a solution of \a.

In this paper we are going to analyze in some detail the physics of \a~for
$N=1$ (the gravitational SG model).  The action is described after
coupling to gravity by:
\eqn\b{
\eqalign{
S=&\int{d^2 z \sqrt{\hat g} \left [ {1\over 8 \pi} (\hat\nabla \phi)^2
+{\mu\over 16 \pi}e^{\sqrt{2} \phi}+{\sqrt{2}\over 4 \pi}\phi R(\hat g)
\right ]}\cr
&+\int{d^2 z \sqrt{\hat g} \left [ {1\over 4 \pi}
(\hat\nabla X)^2
+ \lambda e^{\bp \phi} \cos (pX) \right ]}
}
}
where $\hat g$ is some fiducial metric, $\hat R$ is its curvature,
$\mu$ is the cosmological constant, $\phi$ is the Liouville field,
$\bp = (\sqrt{2}-|p|/\sqrt{2})$, and $X$ is a free scalar field, which we will
take to be compact,  $X\simeq X+2\pi R$,
so that the momenta $p$ are quantized $Rp \in \IZ$.  The non-compact case
is a trivial generalization.
This model was first considered by Moore \Moo~where the spherical
partition sum was calculated as a function of $\lambda$ at fixed
cosmological constant $\mu$.
A physical picture for \b~and in general for all models in
\a~was suggested in \Kut.  According to that picture, for $p<2$ in \b~(so
that the cosine is relevant\foot{$X$ is normalized such that the
dimension of $e^{ikX}$ is ${1 \over 4}k^2$.}), the infrared limit should
correspond
to a set of $n$ decoupled $c=0$, pure gravity models, where $n$ is the
number of minima of the potential in \b~related to $p$ through $p={n
\over R}$.  In this paper we will verify that picture.

The plan is as follows.  In section 2 we use the results of \Moo~for the
spherical partition sum and show that in the infrared the
theory \b~with $p<2$
has $\gamma_{st} = -\half$, the right value for pure gravity.  We also
discuss a finite renormalization of the vacuum energy which arises and
clarify the relation to \Moo.  In section 3 we examine the partition sum
of the model on the torus and find that as expected the
infrared partition sum
is ${n \over 48}$ (for $p = {n \over R}$ in \b ), in agreement with \Kut.
We also show that the torus partition sum is a monotonically decreasing
function of the scale, confirming its interpretation as counting the
number of degrees of freedom
\ref\KS{D. Kutasov and N. Seiberg, Nucl. Phys. {\bf B358} (1991) 600.},
and satisfying a new $c$-theorem \Kut.
In section 4 we briefly describe the fate of the operators in
the ultraviolet $c=1$ model as one flows to the
infrared and the more difficult
case of an irrelevant cosine perturbation
($p>2$ in \b).  Section 5 includes some remarks about our results.  Some
of the details appear in the appendices.

\newsec{The Gravitational SG Model on the Sphere}

Our goal is to calculate the partition sum
\eqn\c{
\CZ_0 = \langle e^{-S(\lambda,\mu)} \rangle
}
where $\langle\cdots\rangle$ denotes a path integral over $\phi$
and $X$ with the
action \b.  Since the theory with $\lambda = 0$ (a scalar field coupled
to gravity) is exactly solvable \matmod, it is natural to try to
calculate $\CZ_0(\mu,\lambda)$ by perturbing in $\lambda$:
\eqn\d{
\CZ_0(\mu,\lambda) = \sum_{n=0}^\infty {\lambda ^n \over n!}
\langle(\cos{pX}~e^{\bp\phi})^n\rangle_{\lambda=0} =
\sum_{n=0}^\infty {(\half \lambda ^2)^n \over (n!)^2}
\langle (e^{ipX+\bp\phi})^n (e^{-ipX+\bp\phi})^n
\rangle_{\lambda=0}
}
In the absence of gravity, this expansion is expected to have a finite
radius of convergence, at least in the super-renormalizable case ($p<2$).
The qualitative argument for that is that the potential
in \a\ is a conformal
primary (or a sum thereof), and does not significantly
alter the large field ($X$) behavior of the action.
Turning on gravity should not spoil this.  Indeed, consider the action
\b.  For $0<p<2$, $0<\bp<\sqrt{2}$ so that the $\lambda$ perturbation is
soft: in the ultraviolet, $\phi \rightarrow -\infty$, the
coefficient of the
cosine $e^{\bp\phi}\rightarrow 0$, whereas in the infrared, $\phi
\rightarrow\infty$ that coefficient is large but it is still a small
perturbation compared to the dominant cosmological term proportional to
$\mu$.  Thus, as in flat space, one expects the expansion \d~to be good
(convergent) for $p<2$ (and any $\lambda$).  For $p>2$ we see
that the $\lambda$
perturbation in \b~qualitatively changes the behavior of the potential
in the ultraviolet region.  Instead of dying off, it now
fluctuates wildly with
an amplitude which diverges as $\phi \rightarrow -\infty$.  Thus a
perturbative expansion in $\lambda$ is expected to be asymptotic in this
case. All of the above discussion is of course independent of topology
and we expect this picture to be valid for all genera.
We will see that these expectations are indeed realized in
the exact solution.

One additional point should be made before we turn to the calculations.
In \b~we have presented the theory at fixed cosmological constant $\mu$.
Although this is not always emphasized in the literature, the more
``fundamental'' quantity in two dimensional gravity is the
fixed area partition sum, in terms of which
\eqn\e{
\CZ(\mu,\lambda)=\int_0^\infty dA~\CZ(A,\lambda)~e^{-\mu A}
}
For example, it is in the fixed area representation that KPZ scaling
holds; the Laplace transform \e~sometimes introduces divergences at
small area which lead to logarithmic scaling violations in $\mu$.
Also, at fixed area the sum over genera is always absolutely convergent
while at fixed $\mu$ is it always asymptotic. Below
we will mention other aspects of this issue, but a more complete
discussion is beyond the scope of this paper.

After these initial remarks, we are ready to present the calculation of
$\CZ_0(A,\lambda)$ \d, \e.  Utilizing the results of \Moo~one finds that
\eqn\f{
A^3\CZ_0(A,\lambda) = \sum_{n=0}^\infty {{\lambda
^{2n}~(1-p)^n~A^{n(2-p)}} \over
{n!~\Gamma (n(1-p)+1)}}
}
In \f~we are using a certain normalization of the tachyon operators
$\CT_p=e^{ipX+\bp\phi}$ which can be controlled by rescaling $\lambda$.
Note that \f~has the structure discussed above: for $p<2$ it is an
absolutely convergent series for all $\lambda$ and $A$, while for $p>2$
it is an asymptotic series, in agreement with the heuristic picture
above.

In the ultraviolet \f~trivially leads to $\CZ_0(A,\lambda) \sim A^{-3}$, as
appropriate for the $c=1$ model coupled to gravity.  It is easy to sum
the series \f~explicitly when it is absolutely convergent ($p<2$).  We
do that in Appendix A with the result (for large $A$):
\eqn\g{
\CZ_0(A) \sim A^{-3-\half} e^{\mu_c (p) A} \;;\;(A\rightarrow\infty)
}
with\foot{$\mu_c$ has also a trivial dependence on $\lambda$
following from
KPZ scaling, $\mu_c(p) \sim \lambda^{2 \over 2-p}$.  Below we set
$\lambda = 1$ with no loss of generality.}
\eqn\h{
\mu_c (p) = {(2-p) \over (1-p)} |1-p|^{2 \over 2-p}.
}
The meaning of the results \g~and \h~is quite transparent given our
introductory remarks.  $\mu_c(p)$ of \g, \h~corresponds to a finite
renormalization of the vacuum energy of the infrared theory compared to that
of the ultraviolet one.
This is of course expected on general grounds since gravity is
sensitive to the vacuum energy of matter (in this case $X$).  The
actual value of $\mu_c$ \h\ is
not universal; e.g. if we perturb the potential in \b\
(as in \a), $\mu_c$ will change.  We could tune it to zero by an
appropriate choice of $\mu$ in \b.  On the other hand the power behavior
$\CZ_0(A)\sim A^{-3-\half}$ of \g~is universal (stable under small
deformations \a) and it implies that in the infrared the theory has
$\gamma_{st} = -\half$.  Since the potential in \b~is real we know that
the theory we get is a unitary matter theory coupled to Liouville; the
only candidate is $c=0$ --- pure gravity, as expected.

A comment on the relation to the work of \Moo~is in order.  In \Moo~the
fixed cosmological constant situation was considered.  The large order
behavior of $\CZ_0(A)$, \g, means that the Laplace transform \e~converges
for $\mu > \mu_c(p)$, with a singularity $\CZ_0(\mu)\sim(\mu -
\mu_c)^\half$ as one approaches $\mu_c$.  Note that $\mu_c(p)>0$ for
$p<1$ and $\mu_c(p)<0$ for $1<p<2$, \h.  For $p<1$ this
means that the infrared fixed point appears
along a line $\mu=\mu_c(p,\lambda)$ while for $1<p<2$, where
$\mu_c$ is negative, this occurs at negative $\mu$ or complex
$\lambda$.  We would describe the situation by saying that the
infrared theory is always pure gravity, and that the only difference between
$p<1$ and $1<p<2$ is the sign of the (non-universal) shift in the vacuum
energy, whose physical significance is unclear to us.

\newsec {The Gravitational SG Model on the Torus}

Further information on the physics of the gravitational SG model can be
obtained by analyzing the partition sum \b~on the torus.  Our interest
in this problem is twofold.  First we would like to verify the
intuitive picture that in the infrared, the model consists of $n$ ($=pR$)
copies of pure gravity ($c=0$) models.  Since the partition sum of a
single pure gravity model is \BK
\eqn\i{
\Omega (A) = A\CZ_1 (A) = {1 \over 48}
}
we expect to find in the infrared limit of \b,
\eqn\j{
\Omega (A) \sim {n \over 48} e^{\mu_c(p) A}
}
with the same $\mu_c$ as on the sphere \h.
$\mu_c$ is expected to be independent of genus because it corresponds
to a bulk effect,
therefore it is insensitive to the topology of the world sheet.

As on the sphere, we expand the partition sum in powers of the interaction
\d, and the problem reduces to the calculation of $\langle\CT_p^n
\CT_{-p}^n\rangle$.
Using the results of \moore, \Moo, \KL\ one finds
\eqn\l{
{\langle \CT_p^n \CT_{-p}^n \rangle}_{h=1} = - R \mu ^{n(p-2)}
(p-1)^n~{n! \over 24}~ (f_n(p) + {g_n(p) \over R^2})
}
where $f_n$ is a calculable polynomial of degree $n+1$ and,
\eqn\m{
g_n(p) = {(-1)}^n {\Gamma (n(2-p)) \over \Gamma (n(1-p)+1)}
}
are related to the spherical results. The polynomials
$f_n(p)$ can be in principle obtained from \moore, and our main technical
problem is to find a compact expression for them for all $n$.
To this end we have calculated the first few $f_n(p)$
(the first nine are reproduced in
Appendix B). On the sphere Moore was able to guess the general form
by studying the roots of analogous polynomials. Here, one finds that
$f_n(p)$ always have $n$ roots with $1\le p\le2$ and one with
$-1\le p\le0$. Unlike the sphere, these roots do not
occur at simple $p$'s in general, but {\it for large n}
the pattern is much simpler. Up to small corrections
we find that
(see Appendix B for more details):
\eqn\n{
f_n(p) = {(-1)^n \over \sqrt{2} n} (\sqrt{\pi n} p +1)
{\Gamma (n(2-p)+\half ) \over \Gamma (n(1-p) + \half )};~~n \rightarrow
\infty
}
In fact, as we show in Appendix B, \n\ is a good approximation for
$f_n(p)$ even for low $n$.
The  overall normalization in \n\ is determined from $f_n(p=0)$ which
is known.
As in \Moo\ we should emphasize that we have only checked this
for low values of $n$, but expect it to be valid in general.

Knowledge of the large $n$ behavior of $f_n(p)$ allows us to
estimate the large area limit of $\CZ_1(A)$. Plugging in \l, \n\ into
\d\ we find\foot{For technical reasons explained in Appendix B the analysis
below is valid for $p<1$ only. While the physics is expected to be similar
for $1<p<2$, a more accurate analysis is needed to reveal it.} as
$A\rightarrow\infty$ (see Appendix B):
\eqn\o{
\CZ_1(A) \sim {p R \over 48 A} e^{\mu_c (p) A}
}
with $\mu_c(p)$ given by \h. This is in agreement with the anticipated
result \Kut, according to which when $p=n/R$ the infrared
theory consists of a decoupled
set of $n$ $c=0$ models. Apart from the by now familiar vacuum renormalization
$\exp(\mu_cA)$, we have $\Omega(IR)\equiv A\CZ_1(A\rightarrow\infty)=n/48$.

A second application of \n\ is to calculate
the function $\Omega(A)\equiv A\CZ_1(A)$ as a function of $A$
to see whether it is monotonically decreasing to its infrared value, as
argued in \Kut. One can use the $f_n(p)$ given in \n\ to calculate
$\Omega(A)$; one might worry that since for small $n$ the difference
between \n\ and the exact result (Appendix B) is not
necessarily small, the small
area behavior of $\Omega(A)$ thus obtained may differ from the correct one.
In practice, however, the difference turns out to be insignificant. We have
checked that by calculating $\Omega$ in two different ways --
using \n\ for all $n$, and using the exact $f_n$ for low $n$
and the approximation \n\ for large $n$'s. The two functions agree to within
a few percent.
In \fig\r{$48\Omega(A)e^{-\mu_cA}$ as a function of $\log A$ for $R=10$
and $n=8\ (p=0.8)$.  The dashed line shows the proper infrared limit
for $n=8$.}\
we show as an example $\Omega(A)$, of course multiplied
by $\exp(-\mu_c A)$ to offset the trivial vacuum energy effects, and also
by $48$ for convenience, versus $\log A$, for $R=10$ and $n=8$ ($p=0.8$).
The resulting function is monotonically decreasing
to the appropriate infrared value, in agreement
with the ideas of \Kut. It is natural to define $\Omega$ as a
``$c$ -- function'' in two dimensional gravity.
In \fig\rr{$48\Omega(A)e^{-\mu_cA}$ as a function of $\log A$ for $R=5$
and $n=1,\ 2,\ 3,\ 4\ (p=0.2,\ 0.4,\ 0.6,\ 0.8)$.
The dashed lines indicate the
different infrared limits, showing the different copies of pure gravity
models.}\ we exhibit more examples of the behavior of this $c$ -- function
for given $R$ and
different values of $n$, for which the infrared theory contains different
numbers of copies of pure gravity models.

\newsec {Other Issues}

The simple picture for the space of theories \a~coupled to gravity which
emerges from the results of \Kut\ and this paper leads of course to many
additional predictions which can in principle be checked against the
matrix model results \moore.  As an example, one may consider the flow of
the operators in the $c=1$ string as one approaches the IR pure gravity
fixed point.  There are three kinds of operators in the UV $c=1$ theory:

{\settabs8\columns
\+&1) Tachyon modes $\CT_p$.\cr
\+&2) Oscillator states (also known as discrete states of ghost
number 2 \ref\Lian{B. H. Lian and G. J. Zuckerman, Phys. Lett. {\bf
B254} (1991) 417, Phys. Lett. {\bf B266} (1991) 21.
}.)\cr
\+&3) ``Ground ring'' or discrete states of ghost number
zero \Lian\ref\Witten{E. Witten, Nucl. Phys. {\bf B373} (1992) 187.}.\cr
}

Consider first the tachyon operators $\CT_p$.  On general grounds one
expects them to flow to the physical operators of the IR minimal models,
which form Kac tables.  In particular, when
the IR theory consists of $n$ copies
of pure gravity, we expect all $\CT_p$ to flow to the $n$ identity
operators in the $n$ decoupled vacua.  This is quite plausible because
of the following argument.  We have observed before that when the
gravitational SG model \b~is slightly perturbed by other cosines, as in
\a, one should find (as $A \rightarrow \infty$)
\eqn\pert{
\CZ (\lambda_i,A) \approx A^{-3-\frac12}e^{\mu_c(\lambda_i,p_i)A }
}
That means that the correlation function of $\CT_{p_i}$ given by
\eqn\newA{
\langle\CT_{p_1}\CT_{p_2}\cdots\CT_{p_n}\rangle\cong{1\over\CZ
(\lambda_i,A)}{\partial\over\partial \lambda_1}
{\partial\over\partial \lambda_2}
\cdots{\partial\over\partial \lambda_n}\CZ(\lambda_i,A)\cong A^n
}
so that all $\CT_{p_i}$ behave in the IR as area operators.  Of course
the conclusion assumes \pert, but that can in principle be checked by
using the general tachyon correlation functions of \moore.  It is also
clear that there are precisely $n$ ($=pR$) distinct area operators,
since from \b~we see that
\eqn\ortho{
\langle\CT_{{l_1}/R}\CT_{-{l_2}/R}\rangle =
\delta_{l_1,l_2}~~~~~l_1,l_2=0,1,2,\cdots,n-1
}
so that $\CT_{l/R}, l=0,1,\cdots,n-1$ are independent whereas
$\CT_{(l+n)/R} \approx \CT_{l/R}$ due to the interaction in \b.
Hence there are precisely $n$ independent area operators in the IR
theory, in agreement with the expected picture.  Of course, if we now
turn on $\lambda_i$ with finite strength, the above picture will be
correct for generic $\lambda_i$, but at multicritical points, higher
minimal models will appear and the spectrum will change appropriately to
accommodate the larger Kac table \Kut.

The ground ring states of the $c=1$ model presumably flow to the ground
ring states of the $c=0$ models.  It is not completely clear what is the
IR limit of the oscillator states.  The only states in the IR fixed
point which are not accounted for are the negative ghost number states
of \Lian\ and in principle the oscillator states could flow to those,
but the details remain to be worked out.  In any case, these issues are
more difficult to study using matrix models since both the discrete
oscillator states and the ground ring operators are far less understood
in that framework than are the tachyon modes.

Another issue we have ignored so far is the physics of \b\ when the
cosine is irrelevant, $p>2$.  In that case we have seen that the fixed
area partition sums are given by asymptotic series both on the sphere
\f, and on the torus \l, \n.  Consider for example the spherical result
\f.  Purely on dimensional grounds, the irrelevant
coupling $\lambda_p$ goes to zero in the IR.  Hence naively the IR
theory is the $c=1$ string.  But this raises a problem.  Since we know \Kut\
that the number of degrees of freedom decreases as we flow from the UV
to the IR, the UV fixed point must have more d.o.f. than the
$c=1$ string, but we know that theories with more d.o.f. than $c=1$ are
tachyonic \KS\ and are not likely to be produced by the flow \b\ with
$p>2$.  Then what is the UV behavior of this model?  To answer that
question one must sum the series \f\ in the region of large effective
coupling.  Since the series is asymptotic one must use resummation
techniques.  One possibility is to Laplace transform
$\CZ(\lambda,A)\rightarrow\CZ(\lambda,\mu)$ as in \e; one finds then as
observed in \Moo\ a series with a finite radius of convergence (in $\mu$,
for $\lambda=1$), converging for $\mu < \mu_c(p)$.  In fact, one can take
$\CZ(\lambda,\mu)$ as fundamental as in \Moo; there is nothing wrong
with that, but from our point of view it corresponds to choosing one of
many possible resummations of \f.
In any case, $\CZ(\lambda,\mu)$ can be easily summed and in the UV,
$\mu\rightarrow\infty$, one finds \Moo
\eqn\uv{
\CZ(\lambda,\mu)\sim\mu^2\log\mu
}
indicative of a $c=1$ behavior at short distances.  This is a
general feature of all resummations of \f.  This
however raises a new problem:  we know that a $c=1$ model in the UV does
{\it not} flow to a $c=1$ model in the IR under the perturbation one
finds in the UV analysis alluded to above.
Different resummations of \f\ give rise in the IR either to the $c=1$ model
or -- via a nonperturbative contribution -- to a collection of
$c=0$ ones. This is in qualitative agreement with the phase diagram
of the flat space Sine- Gordon model, but a few puzzles remain;
a complete understanding of this region in Sine- Gordon coupling
space must await further work.

\newsec {Conclusion}

In this paper we have shown that the matrix model results for the
partition sum of the gravitational Sine-Gordon model lead to a
description of the corresponding RG flow consistent with what one
expects from heuristic considerations.  In particular, the
``$c$ -- theorem''
of \Kut\ for the torus partition sum is satisfied and the number of
d.o.f. in the IR corresponds to a number of decoupled pure gravity
systems.  There are still many interesting open questions. In particular,
the
understanding of the Sine-Gordon model with an irrelevant cosine is
unsatisfactory. A better understanding is certainly needed
for the application of
\b\ to two dimensional string theory.

We have also argued that the KdV and generalized KdV hierarchies of
\matmod\ are hidden in the generating functional of the $c=1$ matrix
model.  There should be a simpler and more elegant way of deriving this
result.  The form of the $c=1$ generating functional obtained in
\ref\DMP{R. Dijkgraaf, G. Moore, and R. Plesser, IAS Preprint
IASSNS-HEP-92/48 (1992).} seems closest to the $c<1$
results and perhaps it can be used to prove our results in much greater
generality and with much less work.

\vskip .2in
\centerline{\bf Acknowledgements}

We would like to thank G. Moore, D. Petrich and S. Kachru for useful
discussions.
This work was partially supported by NSF grant PHY90-21984.

\filbreak
\appendix{A}{Large Area Partition Function on the Sphere}

In this appendix we calculate the large area behavior of \f~for
relevant momenta ($p<2$).  Defining
$\CF(A) = A^3\CZ_0(A)$ and $A^{(2-p)} = x^{(1-p)}$, we have
\eqn\aa{
\CF(x) = \sum_{n=0}^{\infty} {{(1-p)^n x^{n(1-p)}} \over
{n!~\Gamma (n(1-p)+1)}}
}
Laplace transforming \aa\ gives
\eqn\ab{
{\bar\CF}(t) = {\int _0^\infty \CF(x) e^{-xt} dx}
= \sum _{n=0}^\infty {{t^{-1-n(1-p)}~(1-p)^n}\over{n!}} =
{1 \over t} e^{(1-p)~t^{(p-1)}}
}
Consider first $0<p<1$, where large $A$ corresponds to large $x$.
We assume an asymptotic form for $\CF(x)$ at large
$x$, $\CF(x) \sim x^{a(p)} e^{b(p) x^{c(p)}}$.
Substituting this into \ab~and using the method of steepest descent to
evaluate the integral, we
obtain ${\bar\CF}(t)$ for small $t$ in terms of
$a(p)$, $b(p)$, and $c(p)$.  Equating this with the last expression in
\ab, we find
\eqn\ac{
{a(p) = - {1 \over 2}~{(1-p) \over (2-p)}},~{b(p) ={(2-p) \over (1-p)}
{|1-p|}^{2 \over {2-p}}},~{c(p)={(1-p) \over (2-p)}}
}
Transforming variables back from $x$ to $A$, we thus have the desired
large area behavior
\eqn\ad{
\CF(A) \sim {A^{-\half} e^{\mu_c(p) A}}
}
where $\mu_c(p) = b(p)$.

For $1<p<2$, large $A$ corresponds to {\it small} $x$, but in this
case, we may simply
assume $\CF(x) \sim x^{a(p)} e^{b(p) x^{c(p)}}$ for {\it small} $x$.
Performing the same analysis as above, we again
find \ad.

\appendix{B}{Calculations on the Torus}

We use the formula derived in \Moo~for the calculation of the torus
correlation functions and find:
\eqn\ba{
{\langle \CT_p^n \CT_{-p}^n \rangle}_{h=1} = - \mu ^{n(p-2)}
(p-1)^n~{n! \over 24}~ f_n(p)
}
where the $f_n$'s are polynomials of degree $n+1$.
The first nine $f_n$'s are:
\eqn\bb{
\eqalign{
f_1(p) &= p^2 - p - 1\cr
f_2(p) &= 3p^3 - 8p^2 + 3p +3\cr
f_3(p) &= 17p^4 - 72p^3 + 90p^2 - 17p -20\cr
f_4(p) &= 2~(71p^5 - 410p^4 + 842p^3 - 670p^2 + 65p + 105)\cr
f_5(p) &= 1569p^6-11455p^5+32460p^4-43475p^3+24915p^2-1014p-3024\cr
f_6(p) &= 12~(1798p^7-15858p^6+57137p^5-106455p^4+105027p^3-46322p^2+\cr
&~~~~~~~~~~63p+4620)\cr
f_7(p) &= 355081p^8-3669526p^7+16023441p^6-38107279p^5+52657584p^4-\cr
&~~~~~~~~~~40905739p^3+14464814p^2+416424p-1235520\cr
f_8(p) &= 16~(425331p^9-5038436p^8+25830780p^7-74616206p^6+132018894p^5-\cr
&~~~~~~~~~~144550364p^4+92211780p^3-26867634p^2-1440855p+2027025)\cr
f_9(p) &= 9~(16541017p^{10}-220912245p^9+1299983709p^8-4415181390p^7+\cr
&~~~~~~~~~~9504928503p^6-13361715045p^5+12088018191p^4-6518013480p^3+\cr
&~~~~~~~~~~1595066820p^2+120188240p-108908800)
}
}
Numerically, we find the $f_n$'s have $n$ roots between $1<p<2$ of the
approximate form
$C_i = 1+{i-\half \over n}$ where $1 \leq i \leq n$, and one negative
root at
$C_- = -{1 \over \sqrt{\pi n}}$.  To determine how well these
values approximate the exact zeros of the $f_n$'s, $\tilde C_i$ and
$\tilde C_-$, we look at the relative error, $(\tilde C_i - C_i)/\tilde
C_i$.  As shown in \fig\epp{
The relative error of the approximate positive roots of the
polynomials $f_n$ for the different values of $n$.
} and
\fig\epn{The relative error of the approximate negative root of the
polynomials $f_n$ for the different values of $n$.}, where we have
plotted the relative errors of the positive roots and the negative root,
respectively, versus $n$, these approximations get better at larger $n$.
We see from these plots that the relative errors decrease steadily as
$n$ increases, and although this does not constitute a proof, we believe
this behavior will continue at higher values of $n$.  In fact, one may
show that these errors obey some power law,
$(\tilde C_i - C_i)/\tilde
C_i
\simeq\alpha_i
n^{-\beta_i}$, where $\alpha_i < 0.1$, and $\beta_i >
0$.  We may therefore write
\eqn\bc{
f_n(p) = c_n (\sqrt{\pi n} p +1)
{\Gamma (n(2-p)+\half ) \over \Gamma (n(1-p) + \half )}
}
where the $c_n$'s are some $n$-dependent normalization factors.
To fix the $c_n$'s,
we use the fact that
$f_n(0) = g_n(0)$ given by \m~with $p=0$.
Using Stirling's formula ($\Gamma (x) \sim \sqrt{2\pi} x^{x-\half}
e^{-x}$), we find the $c_n$'s to be ${(-1)^n \over \sqrt{2} n}$.
Thus,
\eqn\bd{
f_n(p) = {(-1)^n \over \sqrt{2} n} (\sqrt{\pi n} p +1)
{\Gamma (n(2-p)+\half ) \over \Gamma (n(1-p) + \half )};~~n \rightarrow
\infty
}
Although \bd\ strictly holds only as $n\rightarrow\infty$, it is a very
good approximation for low $n$, and the difference between \bd\ and the
exact $f_n$'s are only a few percent.

The one-loop partition sum for fixed $\mu$ is
\eqn\be{
\CZ_1 = {R \over 24} (1 + {1 \over R^2}) \log\mu + {R \over 24}
\sum_{n \geq 1} {(p-1)^n \over n!} \mu ^{n(p-2)} (f_n(p) + {g_n(p)
\over R^2})
}
We Laplace transform \be\ to fixed area, and
neglect the contribution from the $g_n$'s, since
as follows from Appendix A, it is suppressed by $A^{-\half}$
relative to the contribution of $f_n(p)$ at large area.
Using \bd, we find that the large area partition sum is
\eqn\bg{
\CZ_1(A) \sim {R \over 24A} \sum_{n \geq 1}  {(1-p)^n \over \sqrt{2}n~n!
\Gamma (n(2-p))}
A^{n(2-p)}(\sqrt{\pi n} p +1)
{\Gamma (n(2-p)+\half ) \over \Gamma (n(1-p) + \half )}
}
Using Stirling's formula to expand the gamma functions and approximating
the sum by an integral,
we get
\eqn\bh{
\CZ_1(A) \sim {p R \over 48 A} e^{\mu_c (p) A}
}

Finally, we note that the above analysis for the large area behavior of
the torus partition sum only hold for $p<1$.  For $1<p<2$, the series in
\bg\ alternates in sign, and we can no longer neglect
the error in $\CZ_1$ due to the error in the location of the zeros of
the $f_n$'s.
The point is that $\CZ_1(A)$ is then much smaller as
$A\rightarrow\infty$ than the individual terms in the sum defining it, so that
a small mistake in each term can lead to a large mistake in estimating
$\CZ_1(A)$.

\listrefs
\listfigs
\end